\documentstyle[epsf,aps]{revtex}
\newcommand{\be}{\begin{equation}}
\newcommand{\ee}{\end{equation}}
\newcommand{\bea}{\begin{eqnarray}}
\newcommand{\eea}{\end{eqnarray}}

\begin{document}
\bibliographystyle{prsty}
\draft

\title{
A  Continuum Description of Vibrated Sand
    }

    \author{Jens Eggers$^*$ and  Hermann Riecke$^\dagger$}
    \address{
    $^*$Universit\"at Gesamthochschule Essen, Fachbereich Physik,
    45117 Essen, Germany \\
    $^\dagger$ Department of Engineering Sciences and Applied Mathematics,
    Northwestern University, Evanston, Illinois, 60208, USA \\
}

\maketitle

\begin{abstract}
The motion of a thin layer of granular material on a plate
undergoing sinusoidal vibrations is considered. We develop
equations of motion for the local thickness and the horizontal
velocity of the layer. The driving comes from the violent
impact of the grains on the plate. A linear stability
theory reveals that the waves are excited non-resonantly, 
in contrast to the usual Faraday waves
in liquids. Together with the experimentally
observed continuum scaling, the model suggests a close connection between the neutral
curve and the dispersion relation of the waves, which agrees quite well with
experiments. For strong hysteresis we find localized oscillon solutions.

\end{abstract}
\centerline{\today}


\section{Introduction}
\label{sec:intro}
 
Very little is known about the laws governing the macroscopic motion 
of granular materials, or for short, ``sand''.  Most of our 
information comes from either experiment or microscopic molecular 
dynamics calculations.  An accepted continuum description of sand, 
analogous to the equations of hydrodynamics, is missing.  If such a 
description exists, it would help our understanding enormously, much 
in the same way hydrodynamics has dominated our understanding of 
fluids.

One major difficulty is that sand behaves very differently in 
different flow situations.  If at rest or nearly so, sand behaves like 
a solid, and the packing of particles is very important \cite{JaNa92}, 
while to reach a fluidized state the particles have to be shaken quite 
violently.  In an interesting early paper, Haff \cite{Ha83} deals with 
grains in a nearly compact state.  The opposite limit of low 
densities, where particle interactions are dominated by binary 
collisions, is known as ``rapid granular flow''.  Using methods of 
kinetic theory, this limit has received a great deal of attention 
\cite{JeSa83,SaNe83,LuSa84,JeRi86,WaJa97,SeGo98}.  It leads to complicated 
three-dimensional hydrodynamic equations with non-Newtonian transport 
coefficients that still need to be tested against experiment.

Great interest has been stirred by recent experiments in which a thin 
layer of particles is placed on a plate undergoing sinusoidal 
vibrations 
\cite{FaDo89,DoFa89,MeUm94,MeUm95,LuCl96,AoAk96,ClVa96,MeKn97,LaCl98,BiSh98,BrBi98,ShBi98}. 
 Above a certain vibration amplitude regular and irregular 
surface-wave patterns are excited subharmonically, i.e.  the 
frequency of the waves is half that of the driving.  The observed 
phenomena are very similar to Faraday waves excited in a periodically 
vibrated liquid layer (e.g. \cite{KuGo96a}). 
Typical experimental data are the phase diagram 
of different wave patterns as a function of frequency and 
acceleration, and the wavelength of the patterns as a function of 
driving frequency at fixed acceleration.  The data turn out to be 
independent of container size or shape, so the observed patterns 
represent an intrinsic property of the dynamics of vibrated sand.

The experiments triggered a host of theoretical work in which a wide 
range of approaches has been used: molecular dynamics calculations 
\cite{AoAk96,BiSh98}, simplified particle dynamics \cite{Sh97}, 
semi-continuum theories \cite{Ro98,CeMe97,LaCl98}, phenomenological 
Ginzburg-Landau models \cite{TsAr97}, phenomenological coupled map 
models \cite{VeOt98} and order-parameter models \cite{CrRi98}.  Most 
notably, recent molecular dynamics simulations have reproduced the 
experimental results in great detail \cite{BiSh98}.  The comparisons 
between the continuum-type models and experiments have, however, not 
been very detailed.  Their focus was mostly on reproducing the 
localized excitations of the layer (`oscillons') that have been 
observed experimentally \cite{UmMe96}.  Studies of more general 
order-parameter models \cite{SaBr97b,CrRi98} indicate, however, that 
such localized waves can also arise in non-granular systems and are 
therefore not the hallmark of these systems. In fact, similar
excitations have been observed recently in Faraday experiments with
 shear-thinning clay suspensions \cite{Fipriv}.

The goal of this paper is to come up with a model for waves in 
vibrated sand that is based on physically accessible variables and is 
sufficiently realistic to allow a meaningful comparison and test with 
experiments.  At the same time it should be simple enough to permit 
also analytical investigations, which often provide insights that are 
hard to obtain by numerical means.  Based on the observation that 
the dispersion relation of the excited waves exhibits a continuum 
scaling in the small-frequency regime \cite{Um96,BiSh98,ShBi98}, we propose a 
continuum model.  As indicated above, finding a continuum model based 
on the microscopic behavior of individual grains would be a formidable 
task.  For part of the period, sand rests on the plate in a compacted 
state, while in the remainder of the period it is in free fall, 
leading to a fluidized state.  In addition, little is known about 
boundary conditions near a solid wall \cite{JeRi86,BoMi95}.  Therefore, we 
adopt a purely phenomenological approach that is akin to a
shallow-water theory, i.e.  the vibrated sand is described by its 
height and its horizontal velocity only.  The proposed model differs, 
however, in significant aspects from a fluid-dynamical description.  
In particular, the unvibrated sand layer exhibits no oscillatory 
response.  In contrast to the Faraday waves in liquids, the excited 
waves can therefore not be viewed as arising from the resonant driving 
of damped wave modes of the sand layer.

The paper is organized as follows.
In section \ref{sec:model} we introduce a one-dimensional continuum 
model of vibrated sand.  We discuss the driving mechanism and the 
physical significance of the parameters.  In section \ref{sec:lin} the 
model is linearized around a flat layer, to obtain analytical results 
for the onset of the waves and the dispersion relation between the 
excitation frequency and the wavenumber of the resulting surface 
waves.  The results are compared with recent experiments and 
simulations by Bizon {\em et al.} \cite{BiSh98}.  In the next section 
\ref{sec:nonlin} we turn to the nonlinear behavior of the model.  
First we investigate physical origins for the experimentally observed 
hysteresis in the onset of the waves.  Tuning the parameters in the 
model to give strong hysteresis, we find localized oscillon 
solutions.

\section{The model}
\label{sec:model}
Central to our model is the experimental observation that the position 
of the bottom of the layer of sand is closely modeled by the motion 
of a single, totally inelastic particle \cite{Um96}.  This is because 
all the energy is lost upon impact in the inelastic collisions between 
the grains.  The force driving the patterns is thus proportional to 
the acceleration of the inelastic particle, minus the acceleration of 
gravity $g$.  At each impact, this relative acceleration $\gamma(t)$ 
is strongly peaked and its strength is related to the velocity of 
impact.

We consider very thin layers and assume that they can be characterized 
by their thickness and mean horizontal velocity alone.  For 
simplicity, we will only consider one-dimensional motion.  We will not 
address the nonlinear pattern-selection problem that arises in 
two-dimensional patterns (e.g.  stripes {\em vs.} squares).  From mass 
and momentum conservation considerations we arrive at equations quite 
similar to those of a fluid layer in the lubrication approximation, 
except for some crucial differences to be elaborated below.  The 
equations are
\begin{mathletters}
\label{eq1}
\begin{eqnarray}
&& \bar{h}_{\bar{t}} + (\bar{v}\bar{h})_{\bar{x}} = \left(\bar{D}_1  
\bar{h}_{\bar{x}}\right)_{\bar{x}},
    \label{eq1a}\\
&&\bar{v}_{\bar{t}} + \bar{v}\bar{v}_{\bar{x}} = -\bar{\gamma}(\bar{t})
\frac{\bar{h}_{\bar{x}}}{\sqrt{1+\bar{h}_{\bar{x}}^{2}}}
   - \bar{B}\bar{v} +
\left(\bar{D}_2 \bar{v}_{\bar{x}}\right)_{\bar{x}}. \label{eq1b}
\end{eqnarray}
\end{mathletters}

To contrast all physical quantities from their dimensionless 
counterparts, they carry an overbar.  Equation (\ref{eq1a}) comes from 
mass conservation, with an Edwards-Wilkinson diffusion term on the 
right, which describes the tumbling of grains atop one another 
\cite{EdWi82}.

Equation (\ref{eq1b}) expresses the momentum balance where $\bar{v}$ 
is the horizontal velocity integrated over the layer 
height $\bar{h}$\cite{comment}.  
It contains a driving term proportional to both the acceleration 
$\bar{\gamma}(\bar{t})$ relative to a freely falling reference frame 
as well as the slope $\bar{h}_{\bar{x}}$, since particle motion gets 
started only if the surface is inclined.  The denominator reflects the 
assumption that upon impact the sand grains are isotropically 
dispersed but only those scattered out of the layer contribute to the 
horizontal flux. Note that this ensures finite driving in the limit
$\bar{h}_{\bar{x}}\rightarrow \infty$. Without the denominator we 
numerically found wave solutions which became progressively
higher and more peaked, leading to an unphysical finite-time 
singularity. The second and third term on the right describe the 
internal friction due to vertical gradients in the velocity field
and the viscous-like friction due to horizontal gradients, respectively.  
Since the vertical gradients are not resolved in this thin-layer approach they
lead to a bulk damping term.

 Our model (\ref{eq1}) differs from the fluid problem in a number of 
 ways.  Through the assumption of random scattering of the grains upon 
 impact the driving is nonlinear in $\bar{h}_{\bar{x}}$.  In contrast 
 to liquids the sand layer lifts off the plate when the acceleration 
 of the plate exceeds $g$.  Thus, the acceleration term 
 $\bar{\gamma}(\bar{t})$ vanishes over large parts of the cycle and is 
 largest when the sand layer hits the plate.  In the following we will 
 assume that it consists mainly of a series of $\delta$-functions.  
 There is no surface tension in the granular material.  Instead an 
 additional diffusion term appears in the equation for $\bar{h}$.

It should be emphasized that the effective friction and diffusion 
coefficients $\bar{B}$, $\bar{D}_1$, and $\bar{D}_2$ contain 
implicitly the effect of the solid plate and the varying degree of
fluidization present at different frequencies. It would be quite natural 
to assume that the friction between the sand and the plate 
arises only during the phases when the layer is very close to the plate.
Since the velocity is not continuous through the $\delta$-like impact of the layer on the plate a $\delta$-like friction term is mathematically ill-defined.
For simplicity we therefore smear out the friction over the whole period and 
expect that it can be modeled by an effective value of the 
coefficient $\bar{B}$.
We expect that due to the graininess of the material the 
dissipation due to the vertical gradients will 
increase with decreasing slope of the surface since 
the faster flowing grains near the surface are hindered by the slower grains
underneath. This is not unlike the effect of 
an angle of repose.  The coefficients $\bar{D}_1$ and $\bar{D}_2$ for 
particle diffusion and viscosity are expected to depend on the typical 
velocity of the grains which is related to the impact velocity 
$\bar{v}_{0}$.  They will {\it increase} with the typical velocity of 
the grains, since increasing velocity enhances the diffusive transport 
as well as the momentum exchange due to collisions \cite{Ba54,BiSh98a}.  Thus, 
in general we have 
\be 
\bar{D}_{1,2}=\bar{D}_{1,2}(\bar{v}_{0},\bar{h},\bar{v}), \qquad 
\bar{B}=\bar{B}(\bar{v}_{0},\bar{h},\bar{h}_{\bar{x}},\bar{v}).  
\ee 
Since the experiments are performed over a  small range of 
accelerations the impact velocity is essentially given by the 
frequency.  Thus, the dependence on $\bar{v}_{0}$ implies an apparent 
frequency dependence of the coefficients.  Note, a frequency 
dependence would also arise if an averaging procedure could be applied 
in which the basic equations are averaged over a period of the 
driving.  It should be emphasized, however, that this is not the 
origin of the frequency dependence considered in this paper.

We now make all quantities dimensionless using the filling height 
$\bar{h}_0$ as a length scale and $\bar{\tau}=(\bar{h}_0/g)^{1/2}$ as a time 
scale.  The dimensionless driving $\bar{\gamma}(\bar{t})/g$ depends 
then only on the dimensionless acceleration
\begin{equation}
\label{gamma}
  \Gamma = \frac{A\omega^2}{g}
\end{equation}
of the plate, where $A$ is the amplitude of the harmonic driving with 
frequency $\bar{\omega}$.  The dimensionless dispersion relation has 
then the form $q = q(\omega,\Gamma,\bar{h}_0/\bar{d})$, where 
$\bar{d}$ is the grain diameter.  The remarkable observation from the 
experimental data \cite{Um96,BiSh98,ShBi98} is that in the low-frequency
regime the dispersion relation 
is {\it independent} of $\bar{h}_0/\bar{d}$, i.e.  of particle 
diameter, where $\bar{h}_0/\bar{d}$ varies between 3 and 14.  

\section{Linear theory}
\label{sec:lin}
\begin{figure}[H]
\begin{center}
\leavevmode
\epsfsize=0.4 \textwidth
    \epsffile{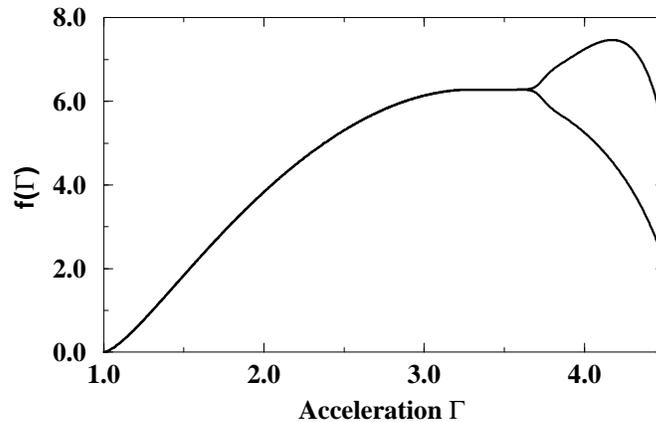}
\vspace*{1.5cm}
\caption{The dimensionless impact velocity of a completely inelastic
ball on vibrating plate as function of the acceleration $\Gamma$.
   }
\label{bounce}
\end{center}
\end{figure}

To understand the instability of the flat layer, we linearize in the 
dimensionless variables $H$ and $V$, with $\bar{h}/\bar{h}_0\equiv h = 
1 + H$ and $\bar{v}/(g\bar{h}_0)^{1/2} \equiv v = V$.  The transport 
coefficients are evaluated at $h=1$, $v=0$.  The linearized equations 
of motion can be transformed into a single wave equation for H:
\begin{eqnarray}
\label{eq7}
H_{tt} = -BH_t + \left(BD_1+\gamma(t)\right)H_{xx} + \nonumber \\
   (D_1+D_2)H_{txx} -D_1D_2 H_{xxxx} .
\end{eqnarray}
To simplify the analytical calculation we assume that $\gamma(t)$
consists only of a series of $\delta$-shocks with period $T=2\pi/\omega$,
\begin{equation}
\label{eq8}
\gamma(t) = v_{0} \sum_{j=-\infty}^{\infty} \delta(t - j T),
\end{equation}
and neglect the force on the layer during the subsequent time intervals
in which the layer is in contact with the plate. Note that $v_0$
is the impact velocity, hence it can be
written as $v_0=f(\Gamma)/\omega$. The curve $f(\Gamma)$ is
shown in Fig. \ref{bounce}, see \cite{Um96}.
As $\Gamma$ rises above 1, the layer
begins to bounce and $v_0$ increases with $\Gamma$.
Between $3.3 < \Gamma < 3.7$ the layer is locked into a state
where it never rests on the plate and $v_0$ remains constant.
Above $\Gamma = 3.7$ period doubling occurs and $\gamma(t)$
can no longer be written in the form (\ref{eq8}). In a straightforward
generalization, two different periods with $T_1 + T_2 = 2T$ appear.

In between shocks, the solution of (\ref{eq7}) is
$H \propto e^{\sigma t} e^{iqx}$, where the dispersion relations
\begin{equation}
\label{eq10}
\sigma_1 = -D_1q^2 , \qquad \sigma_2 = -B - D_2q^2,\label{e:sigma}
\end{equation}
correspond to pure relaxation. At the point of impact,
the $\delta$-function imposes a  jump condition
\begin{equation}
\label{eq9}
H_t^{(+)} - H_t^{(-)} = v_0 H_{xx} .
\end{equation}
The height $H$ itself is continuous.
By making the general ansatz
\[
H_n = \left(\alpha_n e^{\sigma_1(t-nT)} +
      \beta_n e^{\sigma_2(t-nT)}\right)e^{iqx}
\]
and requiring $\alpha_{n+1}=s\alpha_n$ and  $\beta_{n+1}=s\beta_n$ for the
eigenmode we obtain for
the amplification $s$ of the eigenmode
\begin{eqnarray}
\label{eq13}
s = \rho \pm \sqrt{\rho^{2}-s_{1}s_{2}}, \qquad
\rho = \frac{1}{2}\left(s_1 + s_2 -v_0 q^2
  \frac{s_1-s_2}{\sigma_1-\sigma_2}\right) ,
\end{eqnarray}
with $s_1 = e^{\sigma_1 T}$ and $s_2 = e^{\sigma_2 T}$.  Since 
$s_{i}<1$ (cf. eq.(\ref{e:sigma})) and $\rho<1$ the condition for 
instability, $|s|>1$, can only be satisfied with real $s<-1$.  Thus, 
the only instability is subharmonic, i.e.  the motion repeats itself 
only every other period of the driving, as observed experimentally.  
Analyzing (\ref{eq13}) in the limit $q \rightarrow\infty$ shows that 
catastrophic instabilities occur if either $D_1$ or $D_2$ vanish.  
This can also be seen directly from (\ref{eq7}): Since the driving 
with $\gamma$ is proportional to $q^2$, only the combined damping 
$D_1D_2 q^4$ keeps short-wavelength instabilities at bay.  We 
emphasize that this mechanism for wavenumber selection is quite 
different from that at work in the liquid case.  Faraday waves have the 
same dispersion relation as if there was no driving, the significance 
of the driving lies only in exciting the waves.  In the case of sand, 
the medium itself does not support waves as seen from (\ref{eq10}); 
only the competition between  driving at small wavenumbers and damping 
at large wavenumbers selects the wavenumber.

Strictly speaking, (\ref{eq10}) applies only during the free fall 
between shocks.  During the periods in which the sand is in contact 
with the plate and experiences an acceleration it could exhibit 
oscillatory response if the damping is sufficiently weak.  
Specifically, in the absence of any forcing, i.e.  for $\gamma=1$, the 
waves exhibit damped {\it oscillatory} behavior for
\begin{equation}
B(D_2-D_1)<1.\label{e:osc}
\end{equation}
Experimentally, however, the layer spends a large fraction of the 
cycle in free flight already before the onset of waves.  Therefore, 
even if (\ref{e:osc}) should be satisfied during the brief periods 
during which  the layer is in contact with the plate the oscillatory response 
is not expected to be relevant for the excitation of waves.

Next we compute the most unstable mode on the neutral curve
$s=1$. For supercritical transitions
it gives the wavelength that is expected to appear as
the acceleration is raised slowly above the critical value $\Gamma_{c}$.
Introducing the dimensionless combinations
\be
\delta=\frac{{D}_{2}}{{D}_{1}}, \qquad
\beta={B}{T}, \qquad
Q^{2}={D}_{1}{q}^{2}{T},
\ee
we find
\bea
\label{seq1}
{v}_0^{(c)} = {D}_1
\frac{(1+e^{{Q^{2}}})(\beta+(\delta-1)Q^2)}{Q^2}
\frac{1+e^{-(\beta+\delta
Q^{2})}}{1-e^{-(\beta+(\delta-1)Q^{2})}}. \label{e:neutral}
\eea
To find the critical wavenumber, (\ref{e:neutral}) has to
be minimized with respect to $Q$, giving
\be
Q_{c}=Q_{c}(\beta,\delta).
\ee
Plugging this back into (\ref{e:neutral}) leads to the critical impact
velocity $v_0^{(c)}(\beta,\delta)$, and then to an expression of the 
form
\be
f(\Gamma_{c})=v_0^{(c)}\,\omega \equiv D_1 \omega \,\, {\cal F}(\beta,\delta).
\label{e:Gammac}
\ee
In general, the minimum has to be found numerically.
For two limiting cases, however, the
dispersion relation can be given explicitly,
\bea
q_{c} = \frac{0.45}{\sqrt{D_{1}}} \,\, \omega^{1/2}, \qquad \mbox{
for } \beta \rightarrow \infty, \\
q_{c} = a(D_1,D_2) \,\, \omega^{1/2}, \qquad \mbox{
for } \beta \rightarrow 0.
\eea
where $a(D_1,D_2)$ is determined from an implicit equation. In both cases $q_{c}
\propto \omega ^{1/2}$ for fixed $D_{i}$. This is because the waves are
damped by a diffusive mechanism.
The same ``viscous''
scaling has also been found in other approaches \cite{CeMe97,TsAr97}.
Power laws different from $1/2$ are due to some frequency dependence
of the model parameters.

\begin{figure}[H]
\begin{center}
    \leavevmode
\epsfsize=0.4 \textwidth
    \epsffile{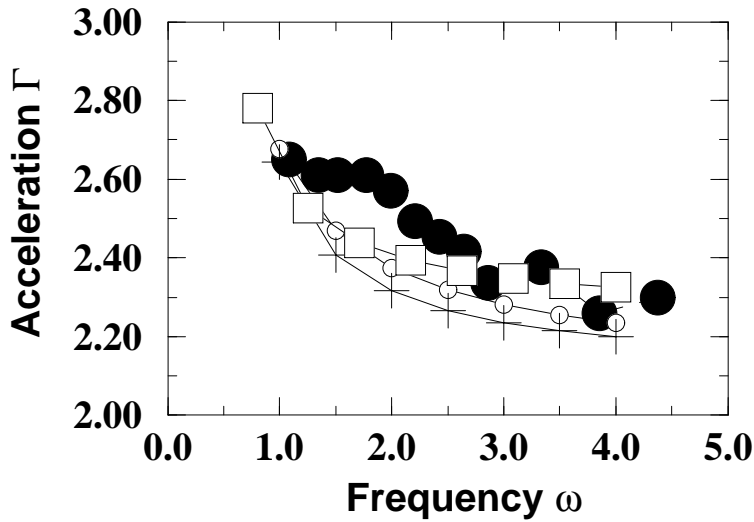}
 \end{center}
\vspace*{1.5cm}
    \caption{
    Neutral curve $\Gamma_c(\omega)$.  Solid circles give the
    experimental results for increasing $\Gamma$ for 
    a mean layer thickness of $2.98 \mbox{mm}$, 
taken from \protect\cite{BiSh98}.
  Theoretical fits for $B=0.08$, $D_1=2.1/\omega$, $D_2=0.12/\omega$ (open squares),
 $B=0.14$, $D_1=0.93/\omega$, $D_2=0.94/\omega$ (plus), and 
$B=0.3$, $D_1=0.1/\omega$, $D_2=1.9/\omega$ (open circles).
   }
       \label{fig2}
 \end{figure}

\begin{figure}[H]
   \begin{center}
       \leavevmode
   \epsfsize=0.4 \textwidth
     \epsffile{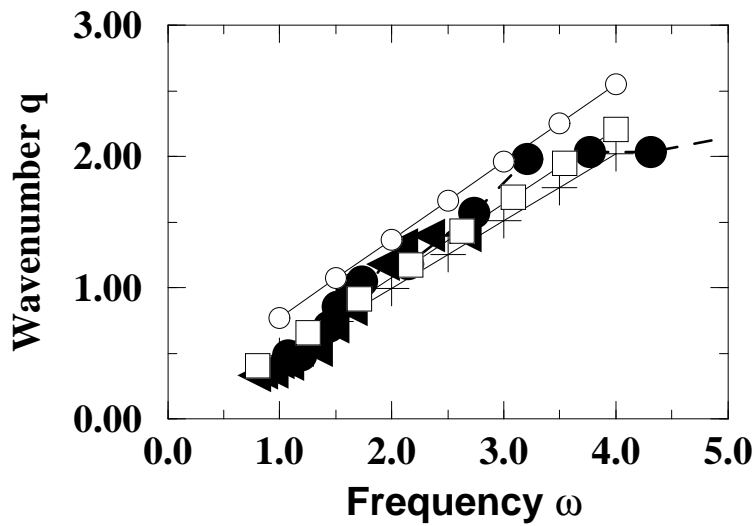}
\end{center}
 \vspace*{1.5cm}
      \caption{
    Dispersion relation $q_{c} (\omega)$
    corresponding to the neutral curves in Fig.\protect{\ref{fig2}}.
Experimental results are given by solid symbols (circles for $\bar{h}_0=2.98mm$, 
triangles for $\bar{h}_0=1.49mm$).
  }
      \label{fig3}
\end{figure}

We can now attempt to make a more quantitative comparison
between experimental measurements \cite{BiSh98} and our model.
We will try to extract the dependence of the transport
coefficients on the layer height and the frequency
from the experimental data, and see whether this leads to a
consistent picture. In Figure \ref{fig2} and \ref{fig3} we show
the experimental measurement of the onset curve and the
dispersion relation, respectively.
In a double-logarithmic plot, the dispersion relation shows a sharp
transition at $\omega = \omega_{tr}$, which lies between 3 and 4 for $\bar{h}_0= 2.98mm$.
In the low-frequency regime $\omega < \omega_{tr}$
the behavior is close to $k \sim \omega^{1.3}$, while a much weaker dependence 
of roughly $k \sim \omega^{0.3}$ is seen at high frequencies.
Note that the exponents of both power laws are much smaller than the exponent
2 reported in earlier experiments \cite{MeUm94}, continuum theory
\cite{CeMe97}, and numerical simulations \cite{AoAk96,LaCl98}.
In view of the short scaling range one should, however, be very careful
interpreting the data in terms of scaling laws.
\vspace*{.4cm}

According to \cite{Um96} the transition between the different
power laws is associated with the frequency
$\Gamma \sqrt{\bar{h}_0/2\bar{d}}$, above which the velocity
of the plate is no longer sufficient to let one particle
hop across the other. Therefore, all particles are locked
into a fixed relative position and the motion is predominantly in the 
vertical direction. Our theory is therefore not expected to be 
applicable. Hence we will only
be concerned with the low frequency regime $\omega < \omega_{tr}$.
From the data of \cite{Um96} for $\bar{h}_0/\bar{d}$ between 3 and 13 it is
also seen that continuum scaling works much better in the
low-frequency regime.

Turning to the phase diagram, stripe patterns are observed
at high frequencies while our frequency
range of interest $\omega < \omega_{tr}$ is associated with
two-dimensional square patterns. This does not, however, affect the comparison
with the {\it linear} properties of the one-dimensional model.
In the low-frequency regime, the critical $\Gamma$ decreases
slightly with frequency, and hysteresis is found.
Experiments at different layer heights \cite{MeKn97,Umpriv}
reveal that there is only a small increase of $\Gamma_{c}$
with layer height.

To compare theory with experiment,
it is useful to return to dimensioned quantities in order to
resurrect the dependence on the mean layer height $\bar{h}_{0}$.
We find
\bea
f(\Gamma_{c})&=&\frac{\bar{D}_1(\bar{\omega},\bar{h}_0)}{g\bar{h}_0}\,
\bar{\omega}\,\, {\cal F}(\beta,\delta), \label{e:neutral2}\\
\bar{q}^2&=&\frac{Q_{c}^2(\beta,\delta)}{2\pi  
\bar{D}_1(\bar{\omega},\bar{h}_0)}\,\,
\bar{\omega}. \label{e:disp}
\eea
Assuming that both diffusion coefficients scale the same way in $\bar{h}_0$ and  
$\bar{\omega}$, we try
the ansatz
\be
\bar{D}_{1,2}=\hat{D}_{1,2}\,\frac{\bar{h}_0^\mu}{\bar{\omega}^\nu}.  
\label{e:dhat1}
\ee
This renders $\delta$ independent of $\bar{\omega}$ and $\bar{h}_0$.
Since experimentally $\Gamma_{c}$ increases only slightly with $\bar{h}_0$ 
at fixed $\bar{\omega}$, we conclude
from (\ref{e:neutral2}) that
$\mu=1+\epsilon$, where $0<\epsilon\ll1$ is used to indicate the weak
increase in $\Gamma_{c}$ with layer height \cite{Umpriv}. This is not
to say that the experimental onset fits such a power-law, 
but rather just to indicate how a change in the dependence of 
the onset on $\bar{h}_0$ affects
the dispersion relation between $q$ and $\omega$
 within the present framework. In all results below
we use $\epsilon=0$. 
Inserting (\ref{e:dhat1}) into the dispersion relation (\ref{e:disp})
one obtains for the
dimensionless wavenumber $q$ as a  function of $\omega$
\be
q^2=\frac{Q_{c}^2(\beta,\delta)}{2\pi\hat{D}_1g^{-\frac{1+\nu}{2}}}\,
\bar{h}_0^{\frac{1-\nu}{2}-\epsilon} \, \omega^{1+\nu}. 
\ee
Experimentally, $q(\omega)$ collapses onto a single curve independent of  
$\bar{h}_0$.
Thus $\nu=1-2\epsilon$ and we obtain
\be
\bar{D}_{1,2}=\hat{D}_{1,2}\frac{\bar{h}_0^{1+\epsilon}}
{\bar{\omega}^{1-2\epsilon}}, \label{e:dhat2}
\ee
where $\hat{D}_{1,2}$ has the dimension of an acceleration (for 
$\epsilon=0$). This makes the
dimension of the material parameters of the sand the same as the control parameter 
of the experiment, which is the physical origin of the observed collapse of the
dispersion relation. Note that (for $\epsilon=0$)
$\bar{D}_{1,2}$ are proportional to the
impact velocity $v_0 \propto \omega^{-1}$ for given acceleration.
Thus, the final result for the  dispersion relation is
\be
q^2=\frac{Q_{c}^2(\beta,\delta)}{2\pi} \, \frac{g^{1-\epsilon}}{\hat{D}_1}\, 
\omega^{2(1-\epsilon)}.
\ee
It remains to determine $\beta$, $\delta$ and $\hat{D}_1$ from a comparison  
with the
experiments.

From (\ref{e:neutral2}) and (\ref{e:dhat2})
it follows that for $B=0$ the onset acceleration $\Gamma_{c}$ is
independent of frequency (for $\epsilon=0$). If $B$ is
taken to be finite, lower frequencies will be damped most,
due to the term $B v$ in (\ref{eq1}). Thus, in accordance with
experimental findings \cite{MeUm95,MeKn97,BiSh98}, the critical
acceleration $\Gamma^{(c)}$ decreases slightly with frequency.

We concentrate on the transition from a flat layer to waves in the
low-frequency regime $\omega < \omega_{tr}$.
The transition occurs for $\Gamma$ between 2.5 and 3 at the
lowest frequencies, depending on the type of particle and on the
layer thickness. For definiteness, we consider the phase diagram
\cite{BiSh98} for a mean layer thickness of $\bar{h}_0 = 2.98 mm$, shown
in Fig.\ref{fig2}. The critical $\Gamma$ is seen to rise from
$\Gamma_c=2.3$ at $\omega=4$ to $\Gamma_c=2.7$ at $\omega=1$.
Keeping the ratio $\delta = D_2/D_1$ constant, one can adjust
$D_2$ to give the correct value for $\omega = 1$.
For $B = 0$, this value would remain constant for all $\omega$.
By choosing $B$ to have $\Gamma_c$ agree with the experimental
finding at $\omega=4$, the onset curve is reasonably well
reproduced. In Figure \ref{fig2} we include the fits for
$\delta=0.057$, $\delta =1.01$, and $\delta=19$,
for which the agreement is comparable.

Next we turn to the dispersion relation of wavenumber versus
frequency as given by Bizon et al. \cite{BiSh98}, shown in Figure \ref{fig3}.
The only parameter remaining to be fixed is the ratio $\delta$ between
damping constants. We adjust it to obtain an optimum fit
of the theoretical dispersion relation to the experimental
data. While the slope of the experimental data is still
slightly higher than predicted from theory, for $\delta =0.057$
we find a reasonable agreement. 

\section{Nonlinear properties and oscillons}
\label{sec:nonlin}

To investigate the nonlinear behavior of the model, the coefficients $D_i$ are taken to 
depend on the local layer height $h$ according to (\ref{e:dhat2}). 
We find that the bifurcation from the flat state to the standing waves
is supercritical if the transport coefficients are taken to be 
independent of $h_{x}$ and $v$ (as in the similar model proposed in \cite{CeMe97}),
while experimentally the transition is subcritical \cite{MeUm94,MeUm95}.
We consider two physically plausible reasons for this 
discrepancy.

Statically, sand has a finite angle of repose. Motion only starts
if the slope of a hill of sand exceeds a certain critical value $\kappa$.
One may expect that in a dynamic state this leads to enhanced
friction of the flowing upper layer when the slope of the surface is
small. We model this by taking
\be
\label{repose}
B=B_{0}\,\left(1+B_{1}\,e^{-(h_{x}/\kappa)^{2}}\right),
\ee
where $\kappa$ sets the characteristic slope below which the granular
character of the material becomes noticeable.

\begin{figure}[H]
  \begin{center}
      \leavevmode
  \epsfsize=0.4 \textwidth
    \epsffile{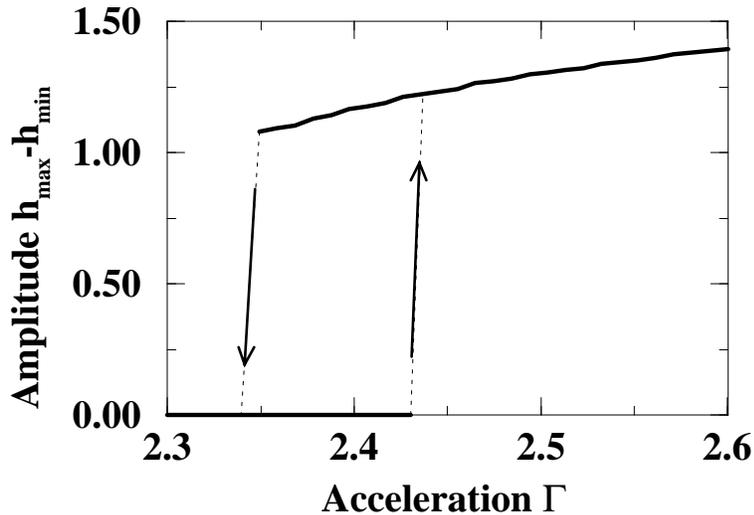}
  \end{center}
\vspace*{1.5cm}
      \caption{
  Hysteretic nature of the transition in the
      presence of a velocity-dependent
     viscosity (\protect{\ref{fluid}}) with
      $\eta=0.5$ and $v_f = 0.2$. The other parameters are $\omega=1.62105$
       $D_1 = D_2 = 0.37$, and $B = 0.45$.
     }
\label{fig4}
  \end{figure}

Second, the layer will have a much higher viscosity when it is near
its compact state. An accurate description will have to contain
a parameter which measures how far the layer is from this
state. As the ``fluidization'' increases, the viscosity is
expected to decrease. The simplest assumption is that the fluidization
depends directly on the local velocity $v(z,t)$, and thus we
model this effect by
\begin{equation}
\label{fluid}
D_2 = D^{(0)}_2(\omega)\left[1 + \eta e^{-(v/v_f)^2}\right] .
\end{equation}
Note that the fluidization and thus the local velocity is distinct from
the typical velocity of the vibrating plate. The latter sets the
typical collision time between particles and thus leads to a
{\it rise} in viscosity as it increases.
It is obvious that  (\ref{repose}) and (\ref{fluid}) can
lead to a subcritical transition to waves. Greater accelerations
are needed to set the layer into motion for zero initial
slope and zero velocity. On the other hand, once waves have
started to appear, the motion will persist to lower values
of $\Gamma$. Fig. \ref{fig4} shows a typical hysteresis with
only the fluidization (\ref{fluid}) taken into account. In the numerical
simulations $\gamma(t)$ is taken to contain not only the 
$\delta$-shocks but also the 
smoothly varying acceleration while the layer is in contact with the plate.

\vspace*{.4cm}

The subcritical properties of the model are believed to play a
crucial role in the appearance of localized subharmonic
excitations called oscillons. They arise from extended square patterns 
when the driving is reduced below the stability limit of the latter.
Alternatively, they can be excited by a localized
``seed'' \cite{UmMe96}.
In one period, an oscillon consists of an axisymmetric jet
of a height several times its diameter shooting out of an
almost flat layer. In the next period, the oscillon forms a
shallow circular trough, surrounded by a small mound.

The appearance of oscillons
requires that the flat layer be linearly stable,
while in the region covered by the oscillon waves can be sustained,
since the grains are fluidized only in the center. 
In the ``trough''-phase, the mound has a very shallow slope facing
out, so the outward flux of material is strongly damped and
 almost all the material is sent radially inward. The small
amount of sand that is transported outward comes back during the
following ``peak''-phase.
Because of mass conservation, it is evident that the formation of 
oscillons is greatly
facilitated by their radial geometry. The greater the radius $r$
away from the center, the smaller the height $h(r)$ that corresponds
to the mass of the peak during the ``peak"-phase. Conversely, ingoing sand is focussed into
a sharp jet. Therefore, within
the one-dimensional version of our model oscillons are expected to arise only
over a smaller range of parameters than in two dimensions \cite{CrRi98}.
In fact, they could only be found for greatly exaggerated
subcritical behavior, i.e. large $B_{1}$ and $\eta$. This may also be 
the reason why so far
stable oscillons have not been found in experiments on vibrated sand between two
narrowly spaced plates, which effectively have only one
horizontal dimension \cite{ClVa96}. In these experiments
localized structures appear only intermittently as localized bursts,
but not as time-periodic structures with steady amplitude.
Our main point will therefore be that structurally our model allows
oscillon solutions. For a more quantitative description an axisymmetric
version of our code has to be considered.

\vspace*{.4cm}
\begin{figure}[H]
  \begin{center}
      \leavevmode
  \epsfsize=0.5 \textwidth
    \epsffile{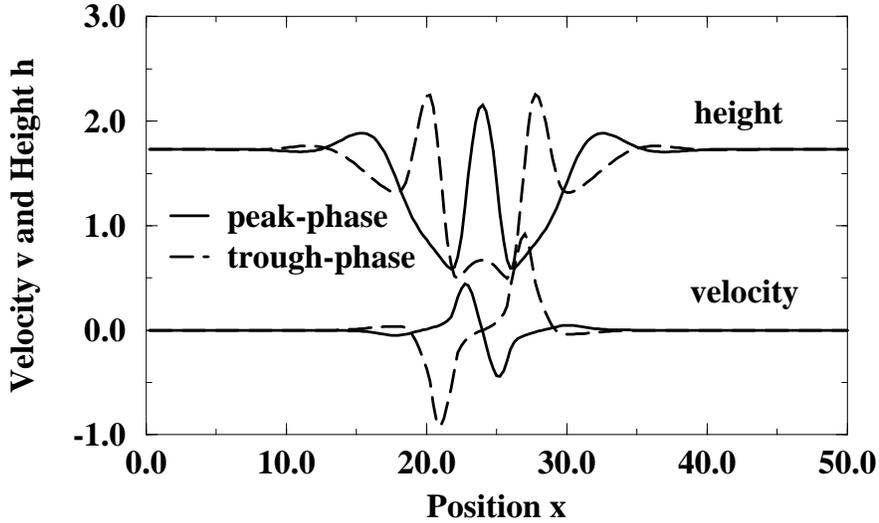}
  \end{center}
\vspace*{1.5cm}
      \caption{
  Oscillon solution for $\Gamma=2.5$, $\omega=1.6$, $D_1=0.376h$, $D_2^{(0)}=0.08h$, 
$B_0=0.1$, $B_1=19$, 
$\kappa=0.2$, $\eta=6$, $v_f=0.4$. 
       }
\label{fig5}
  \end{figure}

\begin{figure}[H]
  \begin{center}
      \leavevmode
  \epsfsize=0.6 \textwidth
    \epsffile{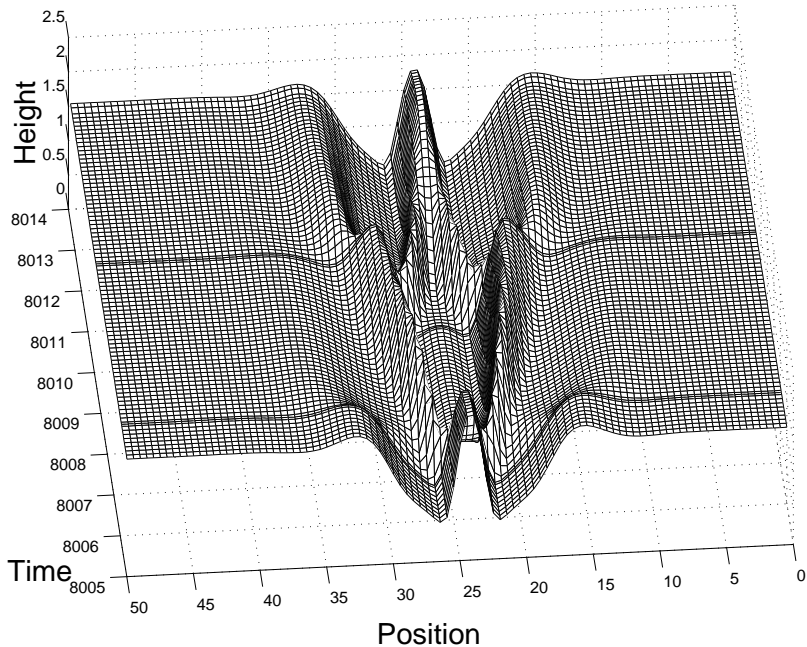}
  \end{center}
      \caption{
   Space-time diagram of the evolution of the oscillon shown in Fig.\protect{\ref{fig5}}
during two
   periods of the driving. The shock-like impact occurs at the times 
   marked with narrower line spacing.
       }
\label{f:oscxt}
  \end{figure}

Figure \ref{fig5} shows a numerically obtained one-dimensional oscillon
in its two phases. The central jet is considerably less sharp
than the experimental one, as expected from the above argument. The parameters 
are chosen so as to damp the motion outside the oscillon very rapidly ($\eta =6$) and to
make sure that as the layer
hits the plate in the trough phase, almost all material is transported
inward ($\kappa=0.2$, $B_1=19$). 
The temporal evolution of the surface height is shown in
Fig.\ref{f:oscxt} as a space-time diagram. It illustrates the
transport away and towards the center of the oscillon.
 As initial conditions,
we chose a localized velocity profile directed inward towards
a point, to produce an initial central peak. We find that the layer
thickness averaged over two periods is smaller inside the oscillon than outside.
Thus, the oscillon tends to push out material. In preparing an initial
condition, we accounted for this by reducing the layer thickness
in the center.
Once the solution was close to stationary, we checked for
exponential convergence towards an oscillon solution.
Furthermore, the parameters could be varied to within a few percent
without destabilizing the oscillon.

\section{Discussion}

In the present paper we have presented a simple 
hydrodynamic description of surface waves on vertically vibrated granular
media. We have 
eliminated all vertical degrees of freedom and describe the 
horizontal motion by three effective transport coefficients.
Our model predicts a new type of instability based on 
the interplay between shocks imparted by impacts on the plate 
and diffusion both in the layer height and the momentum of the 
layer. This distinguishes the model from earlier 
approaches \cite{MeUm94,MeKn97,TsAr97}, which
 in analogy to Faraday waves in liquids are based on the resonant excitation of 
damped waves that
exist even in the absence of periodic driving. 

To obtain quantitative agreement with experiment, the dependence 
of the transport coefficients $B$, $D_1$, and $D_2$ 
on experimental control parameters has
to be taken into account. Remarkably, the possibilities for this dependence 
become severely restricted by the weak dependence of the neutral curve
on the frequency and the collapse 
of the experimental data for the dispersion relation in units of the gravitational
acceleration $g$ and the mean layer height 
$h_0$. As a consequence, the neutral curve and the fact of the data collapse alone
imply a certain dependence of the wavenumber on the frequency, which turns out to 
be in reasonable agreement with the experiment. This leads to  
an interdependence of these seemingly independent 
observations. Conversely, the power laws are
predicted to change if other parameter dependencies of $B$,  $D_1$, and 
$D_2$ apply, in which case data collapse may also no longer occur. 
Thus, the data collapse indicates more than continuum scaling alone, i.e. 
more than the independence of the grain diameter $\bar{d}$;
within our model 
the collapse is due to specific dynamical properties of the vibrated sand
as reflected in the transport coefficients. Indeed, there
are indications \cite{MeKn97,BiSh98} of a high-frequency
regime in which the power law is different and where also the data 
collapse seems to be in question \cite{Um96}.

To obtain the experimentally observed strong sub-criticality of the
transition to waves we have invoked a critical slope $\kappa$ in the friction term (\ref{repose}).
This effect would be connected with the granularity of the material and
the parameters are expected to depend on $\bar{d}/\bar{h}_0$, implying
 that the continuum scaling should
not hold in the nonlinear regime. Unfortunately, the
fluidization (\ref{fluid}) also comes into play when determining the hysteresis,
so there is no simple estimate of the dependence of the hysteresis $\Delta \Gamma$ on
the particle diameter. Still, measurements of $\Delta \Gamma$ as a function of
both $\bar{d}$ and $\bar{h}_0$ would be revealing. 

It is expected that the agreement with experimental data could be further
improved if additional
aspects of the system are included in the model. For instance,
we have neglected effects of a layer dilation, which will
depend on frequency; the dilation will smooth out the $\delta$-driving,
making the driving less effective, and lowering $v_0$.

It would be interesting if the dependence of $B, D_1$, 
and $D_2$ on the frequency or other experimental variables could
be determined directly in experiments. Whether the 
measurement of the frequency dependence of the viscosity
by dragging a sphere through a vibrated layer of sand \cite{ZiSt92}
is directly transferable to our model is not clear.
In addition, it is assumed that only a certain part of the layer takes part
in the horizontal motion. In \cite{CeMe97} this was treated
by introducing a ``penetration depth'', which is an unknown
function of parameters. It was crucial for obtaining a transition
at $\omega_{tr}$ between a high and a low frequency regime \cite{CeMe97}.

An obvious question is whether our model can be tested against 
other experimental observations without significant further
complications and with the same parameter values. 
First, the success \cite{BiSh98} of numerical 
simulations in reproducing experimental data opens the possibility
of a detailed comparison of the wave forms. Nonlinear effects
are very strong and the waves are
far from sinusoidal, the crests being substantially more
peaked than the troughs, which qualitatively is also seen in 
our model. A next step would be an exploration of the phase diagram 
at higher accelerations, at least in the low frequency regime. 
That would necessitate a two dimensional approach to address
the relevant pattern-selection properties.
Such a two-dimensional version would also allow for a quantitative
study of oscillons, both in their shape and their occurrence in 
parameter space.

In the model presented here it is found that localized waves expel material into the 
quiescent regions. Transport of this type 
was considered essential in the Ginzburg-Landau-type
model introduced in \cite{TsAr97}, in which the layer
height couples back to the damping of the waves. 
Such a feedback would occur in the model discussed here for $\epsilon>0$ in (\ref{e:dhat2}). Whether such a transport actually occurs 
in the experiment is of great interest.

A significant step beyond the models presented so far would be an investigation of the 
high-frequency regime where the dispersion relation very 
significantly changes its power law. Since this seems to be
the regime where grains are geometrically obstructing each 
other, one expects a significant dependence on grain
size and a breakdown of continuum scaling. Within our model, 
this could possibly be modeled by a different parameter
dependence of the transport coefficients. On the other hand, 
it is conceivable that the strong spatial correlations introduced 
by the particles, which are locked into a fixed relative position and
predominantly move in a vertical direction, 
calls for a different, perhaps nonlocal hydrodynamic 
description.
 
\acknowledgments
The authors thank H.L. Swinney and P.B. Umbanhowar for
sharing their experimental
data prior to publication.
HR gratefully acknowledges instructive discussions with C. Bizon, M. 
Shattuck, T. Shinbrot, H.L. Swinney, and P. Umbanhowar.
This work was supported by the United States Department of Energy
through grant DE-FG02-92ER14303 and by the Deutsche Forschungsgemeinschaft
trough SFB 237.


\begin{thebibliography}{10}
\bibitem{comment} To obtain
$\bar{h}(\bar{v}_{\bar{t}}+\bar{v}\bar{v}_{\bar{x}})$ in (\ref{eq1b})
from the material derivative $D(\bar{h}\bar{v})/D\bar{t}$
mass conservation (\ref{eq1a}) is used. However, the diffusion
term $(\bar{D}_1 \bar{h}_{\bar{x}})_{\bar{x}}$, which would lead to an  
additional, nonlinear
term in the momentum equation, comes from grains flying above
the layer. This requires a much more sophisticated model to be treated
consistently, and is therefore ignored in the following.
\end{thebibliography}

\end{document}